\documentstyle[12pt]{article}
\begin{document}
\vspace{0.2cm} \centerline{\Large\bf  Effects of density-dependent
quark mass on phase }\centerline{\Large\bf diagram of three-flavor
quark matter } \vspace{0.2cm} \centerline{ Xiao-Bing Zhang and
Xue-Qian Li } \vspace{0.2cm} \centerline{\small Department of
Physics, Nankai University, Tianjin 300071, China} \vspace{8pt}
\vspace{0.2cm}
\begin{minipage}{15cm}
{\noindent   Considering the density dependence of quark mass, we
investigate the phase transition between the ( unpaired ) strange
quark matter and the color-flavor-locked matter, which are
supposed to be two candidates for the ground state of strongly
interacting matter. We find that if the current mass of strange
quark $m_s$ is small, the strange quark matter remains stable
unless the baryon density is very high. If $m_s$ is large, the
phase transition from the strange quark matter to the
color-flavor-locked matter in particular to its gapless phase is
found to be different from the results predicted by previous
works. A complicated phase diagram of three-flavor quark matter is
presented, in which the color-flavor-locked phase region is
suppressed for moderate densities.}

\vspace{0.2cm} {PACS number(s): 25.75.Nq, 12.39.Ba, 12.38.-t}
\vspace{0.2cm}
\end{minipage}

 \vspace{0.5cm} \noindent {\bf I.
INTRODUCTION}\vspace{0.2cm}

The quark matter with three flavors ( $u$, $d$ and $s$ ) has been
intensively studied for two decades. When the down quark chemical
potential is larger than the strange quark mass, the strange quark
matter ( SQM ) might be energetically favored with respect to
two-flavor quark matter and even nuclear matter so that it should
be the ground state of strongly interacting matter \cite{witt}.
Within the framework of the bag model, Farhi and Jaffe pointed out
that SQM with the strange quark mass $m_s <140$MeV ( and with
appropriate bag constant ) becomes the stable ground state for low
baryon densities \cite{jaff}. Based on this consideration, it is
further speculated that some compact stars are made up not of
neutrons but SQM, which are termed as strange quark stars
\cite{ss}. On the other hand, the study of dense quark matter
draws much attention due to the recent progress in understanding
of color superconductivity. At high densities, the original color
and flavor symmetries of three-flavor QCD, namely
$SU(3)_{color}\times SU(3)_{L} \times SU(3)_{R}$, is suggested to
be broken down to a diagonal subgroup $SU(3)_{color+L+R}$ via the
Bardeen-Cooper-Schrieffer ( BCS ) pairing \cite{alf}. The
three-flavor quark matter with the particular symmetry is called
the color-flavor locked ( CFL ) matter and it is different from
SQM which is the matter without the BCS pairing. As another state
of strongly interacting matter, the CFL matter is widely believed
to become " absolutely " stable for sufficiently high densities
\cite{alf03}.

Thus there are two candidates for the ground state of three-flavor
quark matter, CFL and SQM, which are stable for high and low
densities respectively. The question is, in the moderate density
region, which one of them is the ground state. In the other words,
one concerns how SQM undergoes a phase transition to CFL with
increase of density. Investigation on these issues is important
for exploring the physics of strange quark stars and/or the
interior structure of compact stars. Ignoring the $u$ and $d$
quark masses, the CFL free energy takes the form
\begin{eqnarray}
\Omega_{CFL}= - \frac{3\mu^4}{4\pi^2}+ \frac{3m_s^2 \mu^2}{4\pi^2}
-\frac{3\Delta^2 \mu^2}{\pi^2}
 ,\label{pcfl0}
\end{eqnarray}
to order of $m_s^2/\mu$, where $\mu$ is the quark chemical
potential and $\Delta$ denotes the color superconducting gap.
Using Eq.(\ref{pcfl0}),  Alford \emph{et. al.} concluded that the
CFL matter is more stable than the unpaired quark matter (
exactly, SQM ) as long as \cite{alfd}
\begin{equation} {\mu}\geq \frac{m_s^2}{4\Delta}. \label{cflbreak}
\end{equation}
As the necessary condition for CFL existence \cite{alfj02},
Eq.(\ref{cflbreak}) is valid only for high densities. This
inequality can not fully answer the question raised above because
it does not address the phase structure at the moderate densities.

To illustrate this point more clearly, we draw the phase diagram
based on Eq.(\ref{cflbreak}) in the ( $m_s, \mu$ ) plane ( Fig. 1
) \footnotemark[1] \footnotetext[1] { Until now the actual value
of $\Delta$ and its dependence on $\mu$ and $m_s$ are not been
well known, which are closely linked to the gap equation. In the
literature, $\Delta$ was estimated to be of order tens to
$100$MeV. In Fig. 1  it is simply treated as a given parameter
$\Delta \sim 25$MeV for moderate $\mu$, say, $\mu\sim 0.5$GeV
\cite{alf04}.}. When the strange quark mass is small as $m_s
<175$MeV, Fig. 1 shows that SQM is excluded completely from the
moderate density region $\mu=0.3-1$GeV and the CFL matter
including its gapless phase ( gCFL \cite{alf04} ) dominates all
over. However, for small $m_s$, SQM has been predicted to be the
stable ground state \cite{jaff} so that it should be favorable at
least for low densities such as $\mu\sim 0.3$GeV. It implies that
the relation determined by Eq.(\ref{cflbreak}) is problematic more
or less. If assuming that CFL emerges in strange quark stars, this
contradiction becomes more obvious. Starting at very low density
and increasing the matter density by increasing $\mu$, the CFL
formation must be preceded by presence of the stable SQM state.
From this point of view, SQM remains stable for relatively low
densities; otherwise, the self-bound quark stars could not exist
and then the CFL formation would be impossible. Therefore, the
moderate-density phase diagram shown in Fig. 1 needs to be
reexamined, especially in the quark-star environment.

In fact, the implicit assumption for Fig. 1 is that only the
current mass of strange quark $m_s$ was considered in the
descriptions for both CFL and SQM. According to the low-density
QCD, the strange quark mass not merely originates from the
explicit breaking of chiral symmetry. For low densities where SQM
exists as the stable ground state, there is no reason to neglect
the dynamical mass induced by the spontaneous chiral breaking.
Once the dynamical mass is taken into account, it is found that
the SQM stability window, e. g. the allowed region of the current
mass $m_s$, is widened \cite{lug,pion}. This motivates us to
reexamine both SQM and CFL while the dynamical mass is
incorporated. In this work, we will introduce the density
dependence of quark mass to investigate the phenomenological
effects of the dynamical mass on the moderate-density phase
diagram. This approach should be closer to reality and obviously
helpful to clarify the problem of Fig. 1. In Sec.II, we briefly
review the mass-density-dependent model \cite{lug} and consider
the free energy of the CFL matter when the density-dependent quark
mass is incorporated. We emphasize that the mechanism for
existence of the gCFL phase needs to be reexamined in particular
for relatively low densities. In Sec.III, we investigate the phase
transitions from SQM to the conventional CFL phase and/or the gCFL
phase and present a new phase diagram which is very different from
Fig. 1.

\vspace{0.2cm}\noindent {\bf II. THE MODEL }\vspace{0.2cm}

\vspace{0.1cm}\noindent {\bf A. SQM and its stability }
\vspace{0.1cm}

Following Ref.\cite{lug}, the density-dependent quark mass is
given by
\begin{eqnarray} {m}_{D}= {C}/(3\rho),\label{md}
\end{eqnarray}
where $\rho$ denotes the matter density and $C$ is a model
parameter which is constrained by the SQM stability conditions. If
ignoring the current masses of $u$ and $d$ quarks, the masses for
the light and strange quarks in this model are
\begin{eqnarray} {M}_{u}={M}_{d}= {m}_{D};\;\;
{M}_{s}= {m}_{s}+{m}_{D},\label{mq}
\end{eqnarray}
respectively.

The SQM free energy contributed by the Fermi gas reads \cite{lug}
\begin{eqnarray}
{\Omega}(\mu_i,M_i,p^i_F)&=& {{\sum}\atop{i=u,d,s}} \int_0^{p_F^i}
\frac{3}{\pi^2} p^2 (\sqrt{p^2+M_i^2}-\mu_i )dp \nonumber \\
&=& -{{\sum}\atop{i=u,d,s}} \frac{1}{4\pi^2}[{\mu_i}{p_F^i}(
{\mu_i^2}-{5\over2}{M_i^2})+ {3\over2}{M_i^4}\ln
(\frac{\mu_i+p_F^i}{M_i})] .\label{psqm1}
\end{eqnarray}
For each flavor the Fermi momentum ${p_F^i}$ is defined by
$p_F^i=\sqrt{\mu_i^2-M_i^2}$ where $\mu_u=\mu-2\mu_e/3$ and
$\mu_d=\mu_s=\mu+\mu_e/3$ if the electron chemical potential
$\mu_e\neq 0$. On the SQM side, the Fermi momenta of $u$, $d$ and
$s$ quarks are different and are related to the corresponding
densities via $\rho_i={(p_F^i)}^3/{\pi^2}$. Therefore, for SQM,
the electrical neutrality is realized by
\begin{eqnarray}
\frac{2}{3}\rho_u-\frac{1}{3}(\rho_d+\rho_s)=\rho_e=\frac{\mu_e^3}{3\pi^2},
\label{neu}\end{eqnarray} and the baryon density is
\begin{eqnarray} {\rho}&=&\frac{1}{3}(\rho_u+\rho_d+\rho_s).\label{nsqm}
\end{eqnarray}
When the contribution from electrons is included, the total free
energy for SQM becomes
\begin{eqnarray} {\Omega}_{SQM}&=&
{\Omega}(\mu_i,M_i,p_F^i) -\frac{\mu_e^4} {12\pi^2}.\label{psqm}
\end{eqnarray}

With respect to nuclear matter, SQM becomes energetically stable
for low densities as long as its energy per baryon satisfies
\begin{equation} {({\cal E}/\rho)}_{SQM} \le 930\textrm{MeV}, \label{stb}
\end{equation}
at zero pressure, where $930$MeV corresponds to a typical value of
the energy per baryon in nuclei. For our purpose, Eq.(\ref{stb})
needs to be considered seriously to guarantee that not nuclear
matter but SQM undergoes a phase transition to CFL ( if without
this constraint the nuclear-CFL transition \cite{alfd} would be
very likely ). In this work, we fix the parameter $C$ by the
critical condition of Eq.(\ref{stb}) for certainty. For instance,
the value of $C$ is adopted to be $110$ and $70$MeV/fm$^3$ as
$m_s=0$ and $180$MeV respectively ( see Ref.\cite{lug} for details
).

On the other hand, the energy per baryon for two-flavor quark
matter ( 2QM ) is required to satisfy the inequality
\begin{equation} {({\cal E}/\rho)}_{2QM} > 930\textrm{MeV} , \label{1'}\end{equation}
at zero pressure \cite{jaff}. By using Eqs.(\ref{stb}) and
(\ref{1'}), the stability window can be obtained in which SQM
corresponds to the stable ground state at low densities. But
Eq.(\ref{1'}) does not apply to the moderate-density case. Due to
the appearance of strange flavor, SQM is favored over the regular
two-flavor matter as long as
\begin{eqnarray}
\mu_d=\mu+\mu_e/3 \geq M_s.  \label{2s}
\end{eqnarray}
Instead of Eq.(\ref{1'}), thus, Eq.(\ref{2s}) needs to be taken
into account in the following calculation.

\vspace{0.1cm}\noindent {\bf B. CFL and gCFL in the model}
\vspace{0.1cm}

Different from the unpaired one, the CFL quark matter is an
insulator in which no electrons are required for the electrical
neutrality \cite{neu}. The Fermi momenta have the common value
\cite{alfd,neu}
\begin{eqnarray}
\nu=2\mu-\sqrt{\mu^2+m_s^2/3},\label{pfcom}
\end{eqnarray}
for all three flavors and $\mu_e$ does not influence the CFL free
energy directly \footnotemark[2] \footnotetext[2] { Assuming that
the SQM-CFL transition is of first order, there is an interface
between the electron-rich SQM and the electron-free CFL. In this
case, the effective value of electron chemical potential is zero
on the CFL side because of the electrostatic potential at the
metal-insulator boundary ( see Refs.\cite{alfd,zhang} for details
). In the present work, a possibility of the mixed state
consisting of the electrical-opposite SQM and CFL will be ignored.
The reason is that the nonzero $m_D$ provides the additional
instanton interaction so that the electrical-negative phase such
as CFL$K^-$ is very difficult to emerge.}. Thus the CFL free
energy contributed by the Fermi gas is obtained by replacing the
variables $\mu_i$ and $p_F^i$ in Eq.(\ref{psqm1}) by $\mu$ and
$\nu$ respectively. Together with the contribution from the CFL
pairing ( the third term in Eq.(\ref{pcfl0}) ), the total free
energy for CFL takes the form
\begin{eqnarray}
\Omega_{CFL}= {\Omega}(\mu,M_i,\nu)-\frac{3\Delta^2 \mu^2}{\pi^2}
,\label{pcfl}
\end{eqnarray}
when the density dependence of quark mass is considered.

At high density, the value of $m_D$ is close to zero so that the
difference between Eqs.(\ref{pcfl0}) and (\ref{pcfl}) becomes
negligible. For the concerned density region, Eq.(\ref{pcfl})
means that not only $m_s$ but also $m_D$ contribute to the CFL
free energy. As a consequence, the previous results of the SQM-CFL
transition e.g. Fig. 1 might be no longer valid at low/moderate
densities ( see Sec.III for details ).

Now we turn to consider the gapless color-flavor-locked ( gCFL )
phase where the Cooper pairs between the blue-down ( $bd$ ) and
green-strange ( $gs$ ) quasi quarks becomes unstable. At
sufficiently high densities, the strange quark mass is just the
current mass $m_s$ and the light quark masses are very small so
that the mass matrix reduces to $diag(0,0,m_s)$ approximately. It
is well known that the quark mass term can be simplified as
\begin{eqnarray}
-\frac{m_s^2}{2\mu}(\tilde{\psi_L^s}^+ \tilde{\psi_L^s} +
L\rightarrow R), \label{lq}
\end{eqnarray}
at the leading order of the CFL effective Lagrangian
\cite{hong,sch}. Here $\tilde{\psi}$ is not the ordinary quark
field, but the quasi-quark field in the vicinity of the Fermi
surface. For the $bd$-$gs$ Cooper pairs, $\tilde{\psi}_{L/R}^s$ in
Eq.(\ref{lq}) denotes the $gs$ modes near the Fermi surface. It is
worth being notified that Eq.(\ref{lq}) conserves chiral symmetry
although the current mass term in the low-density QCD theory does
not. As suggested in Ref.\cite{sch}, $m_s^2/(2\mu)$ is regarded as
the chemical potential associated with strangeness, i.e.
$\mu_S=m_s^2/(2\mu)$. Obviously, the effective chemical potential
for the $gs$ modes is influenced by the nonzero $\mu_S$ while that
for the $bd$ modes is independent of $\mu_S$. As a result, the
effective chemical potentials $\mu_{gs}^{eff}$ and
$\mu_{bd}^{eff}$ become different. To account for the relative
chemical potential of the paired $bd$ and $gs$ modes, the
variation is \cite{alf04}
\begin{eqnarray}
\delta \mu=\frac{\mu_{bd}^{eff}-\mu_{gs}^{eff}}{2}=
\frac{m_s^2}{2\mu}, \label{dmu0}
\end{eqnarray}
where the contribution from the chemical potential associated with
the color charge has been included. When the variation is larger
than the color superconducting gap, i.e.
\begin{eqnarray}  \frac{m_s^2}{2\mu}\geq \Delta,
\label{gcfl}
\end{eqnarray}
gCFL emerges as the more stable phase than CFL. Based on
Eq.(\ref{gcfl}), gCFL was predicted to emerge at relatively large
$m_s$ and/or relatively small $\mu$, as shown in Fig. 1.

However, this is not the whole story yet. At densities where $m_D$
becomes nonzero, the variation $\delta\mu$ caused by the strange
mass term mainly is influenced by $m_D$ also. This means that the
gCFL existence at moderate densities must be reexamined once the
density dependence of quark mass is considered. Noticing that
$m_D$ is the dynamical mass essentially, it does not enter the CFL
effective Lagrangian via a simple replacement $m_s \rightarrow
M_s=m_s+m_D $ directly. Therefore, an extrapolation like $\delta
\mu \rightarrow \delta
\mu'=\frac{M_s^2}{2\mu}=\frac{(m_s+m_D)^2}{2\mu}$ is not feasible
in principle.

To incorporate the effect of $m_D$ self-consistently, let us
consider the dynamical quark mass term in the Lagrangian involving
the quasi-quark degrees of freedom. Note that the chiral symmetry
pattern exhibits a resemblance to the color-flavor-locking
pattern. It is reasonable to assume that $m_D$ enters the CFL
effective Lagrangian via the chiral-invariant form
\begin{eqnarray}
\xi m_D(\tilde{\psi_L^s}^+ \tilde{\psi_L^s} + L\rightarrow
R).\label{lq2}
\end{eqnarray}
Without losing generality, we introduce an unknown coefficient
$\xi$ in Eq.(\ref{lq2}). If the density-dependent-mass $m_D$ does
account for all the non-perturbative effects of the dynamical
chiral breaking, $\xi$ is equal to $1$ which is adopted in the
calculations of Sec.III. Eq.(\ref{lq2}) is expected to deviate the
strangeness chemical potential from the original value
$\mu_S=m_s^2/(2\mu)$. Correspondingly, only the effective chemical
potential $\mu_{gs}^{eff}$ is influenced by Eq.(\ref{lq2}) for the
$bd$-$gs$ Cooper pairs. In analogy with the treatment of yielding
Eq.(\ref{dmu0}), there exists a decrease in the value of $\delta
\mu$ so that we redefine the variation as
\begin{eqnarray} \delta \mu \rightarrow \delta
\mu'=\frac{m_s^2}{2\mu}-\frac{\xi}{2} m_D.\label{dmu}
\end{eqnarray}

By inserting Eq.(\ref{dmu}) into the dispersion relation for the
$bd$-$gs$ pairs \cite{alfkou}, the gapless modes become possible
at the momenta
\begin{eqnarray}
p^{\pm}_{gapless}= \overline{\mu}\pm \sqrt{(\delta \mu
')^2-\Delta^2},
\end{eqnarray}
where $\overline{\mu}$ is the average value of the $bd$ and $gs$
effective chemical potentials. Compared with the case without
$m_D$, the " blocking " region \cite{liu},
 i.e. the width between $p^{+}_{gapless}$ and $p^{-}_{gapless}$,
is suppressed for not-very-high densities. As a consequence, the
free energy contributed by the gapless phenomenon is expected to
be influenced by both $m_s$ and $m_D$ since the relative free
energy of gCFL and CFL depends on the magnitude of the " blocking
" region mainly.

Based on Eq.(\ref{dmu}), the condition Eq.(\ref{gcfl}) is replaced
by
\begin{eqnarray}\frac{m_s^2}{2\mu}-\frac{\xi}{2} m_D \geq \Delta. \label{gcfl2}
\end{eqnarray}
If CFL emerges at the densities where $m_D$ becomes negligible,
Eq.(\ref{gcfl2}) reduces back to Eq.(\ref{gcfl}) and the critical
condition for the CFL-gCFL transition is still
$m_s^2/(2\mu)=\Delta$ approximately. Also, the necessary condition
Eq.(\ref{cflbreak}) for the existence of the CFL matter needs to
be reexamined in the case of $m_D\neq 0$. Once the variation
between the $bd$ and $gs$ chemical potentials is too large, the
color-flavor-locked pairing might be broken completely. According
to Ref.\cite{liu}, the CFL matter including gCFL exists only when
the variation is not larger than $2\Delta$ which is the energy
cost for breaking a Cooper pair. So the previous condition
Eq.(\ref{cflbreak}) is extended to
\begin{eqnarray}
\frac{m_s^2}{2\mu}-\frac{\xi}{2} m_D \leq
2\Delta.\label{cflbreak2}
\end{eqnarray}
For the moderate density region, Eq.(\ref{cflbreak2}) provides a
more realistic boundary for the CFL matter.

\vspace{0.2cm}\noindent {\bf III. NUMERICAL RESULTS AND
CONCLUSIONS }\vspace{0.2cm}

As a strong-coupling effect, the nonzero $m_D$ is expected to
affect the phase diagram of three-flavor quark matter for
not-very-high densities. Before going to be specific, let us
firstly discuss \emph{whether or not} the phase transition between
SQM and CFL/gCFL occur in the moderate density region. The answer
is not always positive and it is actually linked to the value of
$m_s$, as argued in the following.

Now both SQM and CFL/gCFL are the deconfined phases, therefore the
physics of confinement does not play a role in determining the
SQM-CFL/gCFL transitions. So the negative values of the free
energies obtained in Sec. II are related to the corresponding
pressure directly and then the Gibbs condition for the pressure
equilibrium reads
\begin{eqnarray} P_{CFL/gCFL}-
P_{SQM}={\Omega}_{SQM}-{\Omega}_{CFL/gCFL}=0.
\end{eqnarray}
For $m_s=10$, $50$, $100$ and $150$MeV, we show $\delta
P=P_{CFL}-P_{SQM}$ as a function of $1/\mu$ in Fig. 2. It is found
that $\delta P$ does no longer approach to zero monotonously with
increasing $1/\mu$, i.e. decreasing $\mu$. As shown in Fig. 2,
there exists a rising tendency of $\delta P$ in the vicinity of
$\mu\simeq 0.3$GeV. This leads to the fact that no any pressure
equilibrium appears in the moderate density region so that a
first-order SQM-CFL phase transition does not occur. Although the
CFL pressure is relatively large, the absence of the phase
transition means that the CFL matter is impossible to exist at
least in our concerned density region. Therefore, SQM with small
$m_s$ still remains as a stable state for moderate densities while
CFL with small $m_s$ is prohibited unless the density is very
large \footnotemark[3] \footnotetext[3] { At a very high density,
the pressures for SQM and CFL can become close to each other as
long as the pairing gap $\Delta$ is small enough compared with the
value of $\mu$. In the asymptotic sense, the SQM-CFL transition
always occur regardless of whether $m_s$ is small. But this is not
the case being concerned in the present work. }.

The above physical picture holds valid until $m_s$ is large. For
larger $m_s$, more pressure is paid to maintain the common Fermi
momentum so that the pressure of CFL decreases. As long as $m_s$
is large enough, the SQM-CFL transition in moderate density region
becomes possible to occur. Our numerical calculation shows that,
as $m_s$ is about $150$MeV, the pressure equilibrium comes to
appear in the vicinity of $\mu\simeq 0.3$GeV ( see Fig. 2 also ).
Noticing that such pressure equilibrium behaves like " crossover "
of CFL and SQM, $m_s \simeq150$MeV is regarded as the minimal
value allowed for the CFL existence at moderate density. Once
$m_s$ is larger than $150$MeV, the transition from SQM to CFL/gCFL
might occur in moderate density region.

As a typical example, the result of $\delta P=P_{CFL}-P_{SQM}$ for
$m_s=200$MeV is given by the solid line in Fig. 3. As shown in
Fig. 3, the critical chemical potential $\mu_c$ for the SQM-CFL
transition is about $0.4$GeV. When the gCFL phase and the SQM-gCFL
transition are incorporated, however, the SQM-CFL transition might
not occur in nature since gCFL is more energetically favorable
than CFL. The dashed line in Fig. 3 gives the value of
$P_{gCFL}-P_{SQM}$ for $m_s=200$MeV. As shown in Fig. 3, the
SQM-gCFL transition occurs at the critical chemical potential
$\mu_c'\simeq 0.35$GeV, which is smaller than the value of
$\mu_c$. Therefore, it is not CFL but gCFL to emerge firstly in
the quark star environment. Also, the critical point for the
gCFL-CFL transition $\mu_c''$ is shown to be about $0.67$GeV in
Fig. 3. For $m_s=200$MeV, we conclude that, gCFL exists as a
stable state in the region of $\mu_c'<\mu \leq \mu_c''$ and CFL
emerges as $\mu>\mu_c''$. Here we would like to emphasize the
importance of the nonzero $m_D$ for the properties of gCFL. In
Eq.(\ref{dmu}) the $m_s^2/(2\mu)$ and $m_D$ terms provide opposite
contributions to the gCFL pressure actually. Within the framework
where the density dependence of quark mass is considered, if the
gapless phenomenon were determined by $m_s^2/(2\mu)$ exclusively
the pressure of gCFL would be much larger than that of SQM with
increasing $1/\mu$, as shown by the dotted line in Fig. 3. In that
case, the SQM-gCFL pressure equilibrium and thus the corresponding
transition no longer exists in the moderate density region.
Therefore the effect of $m_D$ is relevant to the gCFL presence for
moderate densities.

Based on the above arguments, a schematic phase diagram is given
for the moderate density region in Fig. 4. There are three kinds
of different structures in the phase diagram according to the
value of $m_s$ :

(i) As $m_s$ is small such as $m_s <150$MeV , the effect of $m_D$
prohibits the CFL formation for not-very-high densities. In this
case, it is not CFL but SQM to be the stable phase in the whole
moderate density region, as shown in Fig. 4. This conclusion is
very different from that obtained by Fig. 1, but agrees with the
original prediction that CFL with zero ( or small ) $m_s$ becomes
possible only when the density is high enough \cite{alf}. When
$m_s$ is slightly larger than $150$MeV, a first-order transition
from SQM to CFL takes place. For instance, let us consider the
horizontal line of $m_s=175$MeV in Fig. 4. At the critical
chemical potential $\mu_c \simeq 0.35$GeV, the SQM-CFL transition
occurs so that SQM remains stable for $\mu<\mu_c$ whereas CFL
emerges for $\mu\ge\mu_c$.

(ii) As $m_s$ is large, the gCFL phase becomes more likely than
the conventional CFL phase and then the SQM-gCFL transition
replaces the SQM-CFL transition to be relevant to the phase
diagram. Considering the horizontal line of $m_s=200$MeV, gCFL
comes to emerge at $\mu_c'\simeq 0.35$GeV and it undergoes a
continuous transition to CFL at $\mu_c''\simeq 0.67$GeV.
Interestingly, the SQM-gCFL transition curve in the low density
region is qualitatively different from the result of Fig. 1 : the
critical value of $m_s$ does increase with decreasing $\mu$, as
shown by the solid line in Fig. 4. The reason is that the gapless
phase is determined by Eq.(\ref{dmu}), in which the $m_s^2/(2\mu)$
and $m_D$ terms offer opposite effects on the gCFL pressure. For
lower $\mu$ the latter effect is more important, so that the
transition becomes possible only when $m_s$ is relatively large.
Also, the continuous gCFL-CFL transition curve is given by
Eq.(\ref{gcfl2}) as shown by the dashed line in Fig. 4. Together
with the SQM-CFL and SQM-gCFL transition curves, there exists an
intersection of the three phase transitions in the vicinity of (
$m_s$,$\mu$ )= ( 185MeV,0.4GeV ).

(iii) As $m_s$ is larger, the variation $\delta \mu'$ might be too
large to allow existence of the color-flavor-locked pairing in
quark matter. Most importantly, the necessary condition
Eq.(\ref{cflbreak2}) for existence of the CFL matter no longer
coincides with the SQM-gCFL transition curve, although
Eq.(\ref{cflbreak}) does in the case without $m_D$. Thus, the gCFL
phase region is surrounded by the CFL boundary curve ( that
obtained from Eq.(\ref{cflbreak2}) ), the SQM-gCFL transition
curve ( that determined by the pressure equilibrium ) and the
gCFL-CFL transition curve ( that obtained from Eq.(\ref{gcfl2}) ),
as shown in Fig. 4. Comparing with Fig. 1, we find that the gCFL
phase region is suppressed for low $\mu$. On the other hand, the
existence of three-flavor quark matter might be prohibited if
$m_s$ is very large. The boundary curve of SQM is given by
Eq.(\ref{2s}) and is shown in Fig. 4 , which seems to imply that
2QM is irrelevant to the gCFL presence. At this point, it must be
stressed that a possibility of the 2QM-gCFL transition could not
be ruled out simply if the color superconductivity exists in the
two-flavor quark matter. In that case, the boundary of SQM shown
in Fig. 4 needs to be modified also. Details involving such 2QM
and its transition to gCFL and/or SQM depend on what kind of
two-flavor color superconducting phase be taken into account,
which is beyond the scope of the present work.

In summary, we extend the descriptions of CFL and gCFL from
high-density case to the moderate density region when the density
dependence of quark mass is considered. Starting at low density
and raising the matter density, the physical picture that SQM
remains as the stable state at first and then undergoes a
first-order phase transition to gCFL and/or CFL is examined in
details. As a result, we predict a more complicated phase diagram
of three-flavor quark matter, in which the CFL/gCFL phase region
is suppressed for low densities. The present phase diagram is
helpful to better understand the ground state of three-flavor
quark matter in the environment of quark stars. Of course, there
are some uncertainties of the color superconducting gap used in
this work. When the value of the gap is chosen in other ways, we
can give the similar phase diagram as Fig. 4. For instance, if the
gap is large such as $\Delta\sim 80$MeV \cite{kap} we find that
the minimum value of $m_s$ allowed for the CFL existence increases
so that the CFL phase region for moderate densities might be
further suppressed. Even if the density dependence of the gap is
included, the change of $\Delta$ in the finite region of
$\mu=0.3-1$GeV is not too drastic and the conclusion obtained from
Fig. 4 is still qualitatively correct. In the further work one
should construct the dynamical quark mass within a more realistic
framework such as that beyond the bag model as well as take the
contributions from the color-sextet pairing and the gap equation
into account. Some of the problems are being investigated.

\vspace{0.5cm} \noindent {\bf Acknowledgements} \vspace{0.5cm}

This work was supported by National Natural Science Foundation of
China ( NSFC ) under Contract No.10405012.

\newpage
\begin{figure}
\caption{Schematic phase diagram in the ( $m_s$,$\mu$ ) plane,
where the solid line is the boundary of the CFL matter obtained
from Eq.(\ref{cflbreak}) and the dashed line is the phase
transition from the conventional CFL phase to the gCFL phase
obtained from Eq.(\ref{gcfl}) ( see the following ).}\label{1}
\end{figure}

\begin{figure}
\caption{The CFL pressure vs. the SQM pressure. The solid lines
from top to bottom are the relative pressures for $m_s=10$, $50$,
$100$ and $150$MeV respectively.}\label{2}
\end{figure}

\begin{figure}
\caption{The CFL/gCFL pressures vs. the SQM pressure for
$m_s=200$MeV, where $\xi$ is chosen to be 1 for simplicity. The
solid ( dashed ) line is the relative pressures for CFL ( gCFL ),
while the dotted line is that for gCFL which is determined by
Eq.(\ref{dmu0}) exclusively.}\label{3}
\end{figure}

\begin{figure}
 \caption{ Similar as Fig. 1 but the SQM-CFL/gCFL transitions are
considered in the case of including effects of $m_D$. The solid
lines are the first order SQM-CFL/gCFL transitions, the dashed
line is the continuous CFL-gCFL transition and the dot-dashed
lines correspond to the boundaries of two kinds of three-flavor
quark matter.}\label{4}
\end{figure}

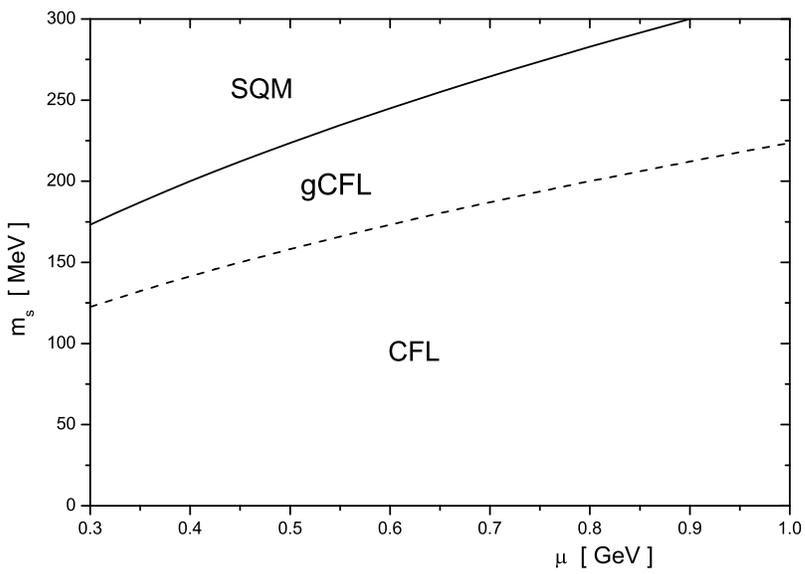
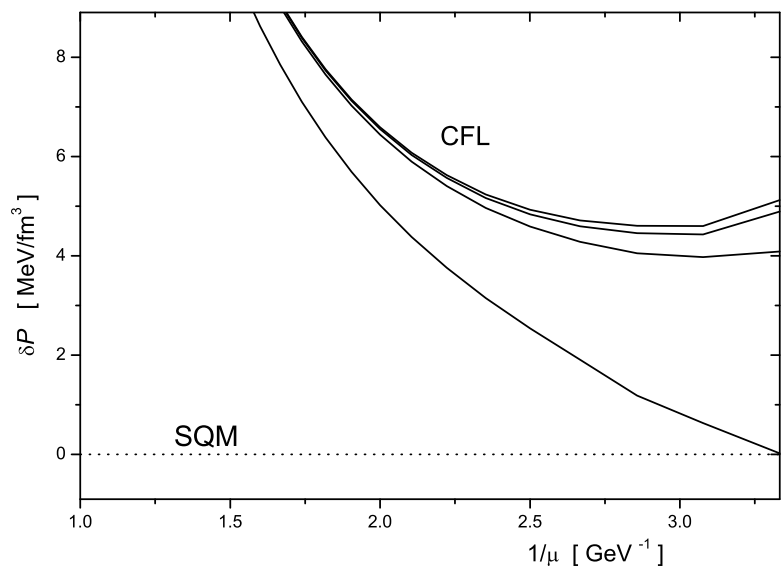
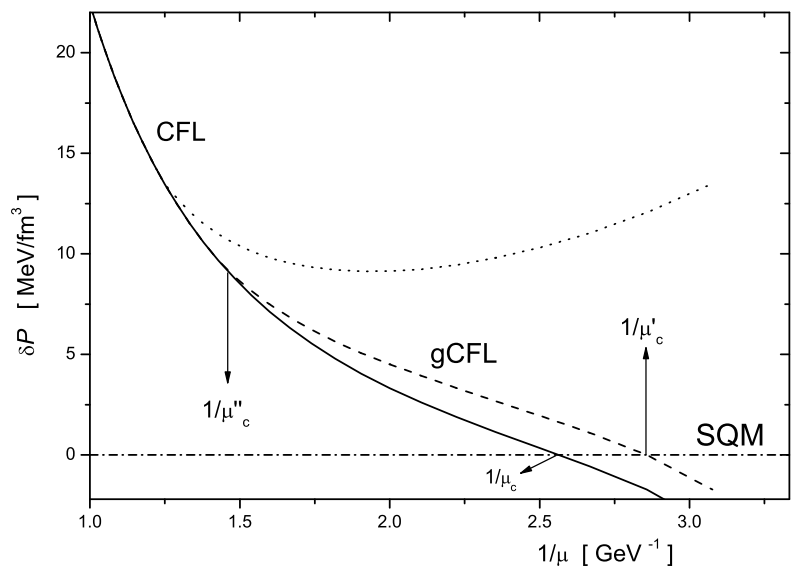
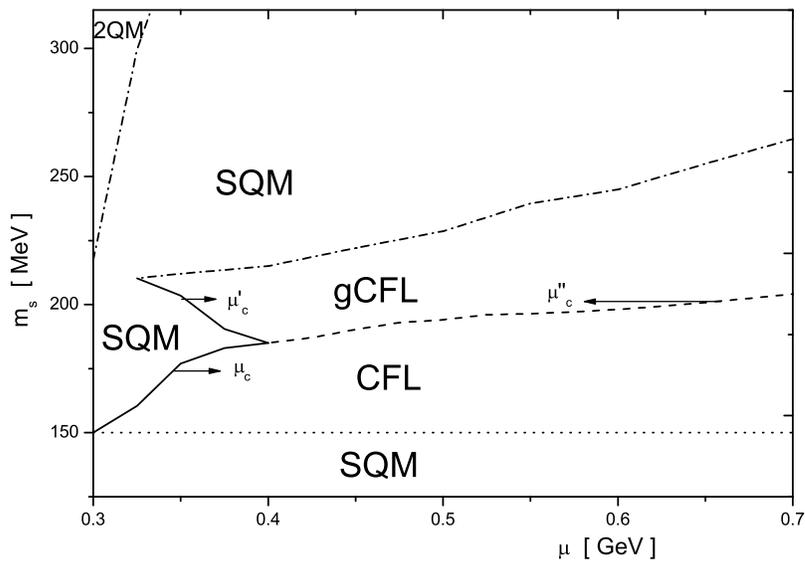
\end{document}